\documentclass[prl,aps,twocolumn,showpacs,superscriptaddress,amsfonts,amsmath,floatfix]{revtex4}

\usepackage{graphicx}
\usepackage{color}

\begin{document}


\title{Quasimomentum distribution and expansion of an anyonic gas}

\author{Tena Dub\v{c}ek}
\affiliation{Department of Physics, Faculty of Science, University of Zagreb, Bijeni\v{c}ka c. 32, 10000 Zagreb, Croatia}

\author{Bruno Klajn}
\affiliation{Department of Physics, Faculty of Science, University of Zagreb, Bijeni\v{c}ka c. 32, 10000 Zagreb, Croatia}

\author{Robert Pezer}
\affiliation{Faculty of Metallurgy, University of Zagreb, Aleja narodnih heroja 3, 44103 Sisak, Croatia}


\author{Hrvoje Buljan}
\email{hbuljan@phy.hr}
\affiliation{Department of Physics, Faculty of Science, University of Zagreb, Bijeni\v{c}ka c. 32, 10000 Zagreb, Croatia}

\author{Dario Juki\'{c}}
\affiliation{Faculty of Civil Engineering, University of Zagreb, A. Ka\v{c}i\'{c}a  Mio\v{s}i\'{c}a 26, 10000 Zagreb, Croatia}

\date{\today}

\begin{abstract}
We point out that the momentum distribution is not a proper observable for a system of anyons in two-dimensions.
In view of anyons as Wilczek's composite charged flux-tubes, this is a consequence of the fact that the orthogonal components of the kinetic momentum operator do not commute at the position of a flux tube, and thus cannot be diagonalized in the same basis. As a substitute for the momentum distribution of an anyonic (spatially localized) state, we propose to use the asymptotic single-particle density after expansion of anyons in free space from the state.
This definition is identical with the standard one when the statistical parameter approaches that for bosons or fermions. Exact examples of expansion dynamics, which underpin our proposal, and observables that can be used to measure anyonic statistics, are shown.
\end{abstract}

\maketitle

Anyons are quantum particles residing in two dimensions (2D), obeying fractional statistics interpolating between bosons and fermions~\cite{Wilczek1982, Leinaas1977}. The only physical realization of anyons so far is found in the fractional quantum Hall effect (FQHE)~\cite{Tsui1982,Laughlin1983}, where localized quasiparticle excitations have a fractional elementary charge~\cite{Laughlin1983} and statistics~\cite{Arovas1984,Halperin1984}. 
While fundamental motivation for exploring anyons is self-evident, the so-called non-Abelian anyonic excitations hold potential for technological advances, as they could be used for robust topological quantum computation~\cite{Kitaev2003} (for review see Ref.~\cite{Nayak2008}).

Some of the intriguing quantum mechanical implications of fractional statistics were pointed out decades ago~\cite{Wilczek1982, Leinaas1977}.
Experiments with ultracold atomic gases seem to be a perfect playground for exploring anyonic physics,
because of the quality in preparation, manipulation, and detection of numerous intriguing quantum states~\cite{Bloch2008}, and because of the possibility to explore 2D systems~{\cite{Hadzibabic2006}, with synthetic magnetic fields ~\cite{Lin2016}, which could be used to tinker with statistics. In an early paper, Paredes {\it et al.}, inspired by the FQHE, proposed the realization of a $1/2$-Laughlin state in a bulk rotating gas~\cite{Paredes2001}. Different schemes were later proposed with atoms in optical lattices~\cite{Duan2003,Aguado2008,Zhang2014}. Ultracold atoms with two hyperfine levels in non-Abelian potentials could yield ground states with non-Abelian anyonic excitations~\cite{Burrello2010}, while bosons in Floquet-driven optical lattices may effectively exhibit fermionic statistics~\cite{Sedrakyan2015}.
The one-dimensional (1D) version of anyons~\cite{Batchelor2006,Patu2007,Santachiara2008,delCampo2008,Hao2008,Keilmann2011,Tang2015, Greschner2015,Strater2016} has also aroused interest, especially in 1D optical lattices~\cite{Keilmann2011,Tang2015,Greschner2015,Strater2016}.
Such particles were proposed to emerge from occupation dependent hopping amplitudes, which could be realized with laser-assisted tunneling~\cite{Keilmann2011,Greschner2015}, or Floquet modulations~\cite{Strater2016}. Other proposals include lattices of polar molecules~\cite{Micheli2006}, photonics lattices~\cite{Longhi2012} and circuit-QED systems~\cite{Kapit2014}.
An undoubtedly important ingredient that needs to be investigated in this context is the detection of the anyonic quantum state. The studied detection schemes rely on braiding~\cite{Paredes2001, Aguado2008, Kapit2012, Grass2014}, the pair-correlation function~\cite{Zhang2014}, and precision spectroscopy~\cite{Cooper2015}.
Free expansion, or time-of-flight method, is among the most used detection techniques from the atomic physics toolbox~\cite{Bloch2008}, which could be of interest for systems where bosons are converted into anyons by manipulating with their interaction (as in~\cite{Keilmann2011}), or by introducing them as impurities in a background of topological states (e.g.,~\cite{Aguado2008,Zhang2014,Lundholm2016}). We are aware of expansion studies only in 1D systems~\cite{delCampo2008}, but not for 2D anyons.

Here we study the expansion of (Abelian) anyons in 2D space. For a system of ultracold bosons or fermions, free expansion provides the momentum distribution of the initial quantum state~\cite{Bloch2008}, which is defined as the diagonal of the reduced single-particle density matrix (RSPDM) represented in a basis of kinetic momentum eigenstates.
The momentum distribution was of paramount importance as a signature of Bose-Einstein condensation~\cite{Cornell2002}, and the onset of Fermi degeneracy in a trapped atomic gas~\cite{DeMarco1999}.
First we point out that the momentum distribution is not a proper observable for anyons. If we think of anyons as charged flux tubes~\cite{Wilczek1982}, then this follows from the fact that orthogonal components of the kinetic momentum operator do not commute at the position of a flux tube, and cannot be diagonalized in the same basis.
However, this can be remedied by turning definitions around: we define the quasimomentum distribution for anyons as the asymptotic limit of the single-particle density of an anyonic gas freely expanding from an initially localized state, which reduces to the standard definition in the case of bosons or fermions.
As an example, we calculate an exact time-dependent wavefunction which for $t{<}0$ describes an eigenstate of $N$ anyons in a harmonic trap, and for $t{>}0$ describes expansion of anyons after the trap is suddenly turned off at $t{=}0$. The solution is found by employing a scaling transformation, and the quasimomentum distribution via Monte-Carlo integration.
For $N{=}2$ particles, we find that the asymptotic single-particle density corresponds to the projection coefficients of the initial state onto two-anyon eigenstates in free space, which underpins our conjecture.
In addition, we point out that anyonic statistics can be extracted from the pair-correlation function: the two-particle correlations at short interparticle distances scale as a power-law with the statistical parameter $\alpha$ in the exponent.

An anyonic wavefunction $\psi$ describing expansion from an eigenstate in a harmonic trap obeys the Schr\"odinger's equation
$i \frac{\partial}{\partial t}\psi{=}H\psi$, with the Hamiltonian
\begin{equation}
\label{H}
H=\sum_{i=1}^N \left[
-\frac{1}{2}{\bf\nabla}_i^2
+\frac{1}{2}\omega(t)^2 r_i^2
\right].
\end{equation}
Here, $\omega(t{<}0){=}1$ and $\omega(t{\geq}0){=}0$.
The symmetry of the wavefunction is anyonic, i.e. $\psi(\ldots,{\bf r}_i,\ldots,{\bf r}_j,\ldots,t){=}e^{i m\pi\alpha} \psi(\ldots,{\bf r}_j,\ldots,{\bf r}_i,\ldots,t)$, where ${\bf r}_i {=} x_i {\bf \hat x}{+}y_i {\bf \hat y}$
are the particle positions, and $m {\in} \mathbb{Z}$ depends on how they are braided during the exchange.
The anyonic wavefunction is a multi-valued function of the positions $\{{\bf r}_i\}$ (e.g., see~\cite{Halperin1984} for a discussion).
For bosons and fermions, the RSPDM,
$
\rho({\bf r},{\bf r}',t)
{=}{N}{\int}{\psi^*({\bf r},{\bf r}_2,...,{\bf r}_N,t) \psi({\bf r}',{\bf r}_2,...,{\bf r}_N,t)} d{{\bf r}_2}...d{{\bf r}_N} ,
$
furnishes one-body observables such as the momentum distribution, which is given by its Fourier transform:
$n({\bf k}{,}t){=}(2 \pi)^{-2} \int \rho({\bf r},{\bf r}',t) e^{i {\bf k}{\cdot }({\bf r}-{\bf r}')} d{\bf r}d{\bf r}'$.
For anyons, the single-particle density, i.e., the diagonal $\rho({\bf r},{\bf r},t)$ of the RSPDM is uniquely defined, as it is not phase-dependent; therefore, $\rho({\bf r},t){\equiv}\rho({\bf r},{\bf r},t)$ is a legitimate observable.
However, the off-diagonal elements of the anyonic RSPDM depend on the wavefunction phase and are not single-valued. Consequently, $n({\bf k},t)$ is not single-valued and therefore it cannot be used as a definition of momentum distribution for anyons.
We note in passing that for 1D anyons this problem does not exist as the wavefunction and consequently the RSPDM are single valued~\cite{Patu2007,Santachiara2008,delCampo2008,Hao2008,Keilmann2011}.

A more physical insight in the question of anyonic momentum distribution is obtained
if we think of anyons as Wilczek's composite particles (CP) consisting of a point charge $q$ and an infinitely thin magnetic flux tube with magnetic flux $\Phi$, so that $\alpha{=}{-}q \Phi{/}2 \pi$~\cite{Wilczek1982}.
The Hamiltonian describing such composite particles includes pairwise vector potential interactions:
\begin{equation}
\label{H_{CP}}
H_{CP}= \sum_{i=1}^N \left[
-\frac{1}{2}\left({\bf \nabla}_i +i \alpha \sum_{j \neq i}
\frac{ {\bf \hat z} \times {\bf r}_{ij}  }{r_{ij}^2} \right)^2
+  \frac{1}{2}\omega^2(t) r_i^2 \right],
\end{equation}
where ${\bf r}_{ij}{=}{\bf r}_i{-}{\bf r}_j$.
The corresponding wavefunction $\psi_{CP} ({\bf r}_1,\ldots,{\bf r}_N,t)$ is bosonic or fermionic (here we assume bosonic symmetry for $\psi_{CP}$). The vector potential interactions
can be gauged out from the Hamiltonian $H_{CP}$ to obtain $H$ \cite{Wilczek1982,Wu1984},
that is, the wavefunction $\psi_{CP}$ is related to the anyonic wavefunction $\psi$
by a gauge transformation
\begin{equation}
\psi \left({\bf r}_1,...,{\bf r}_N,t\right)
=\prod_{i<j}^N e^{i \alpha \phi_{ij}}
\psi_{CP} \left({\bf r}_1,...,{\bf r}_N,t\right),
\end{equation}
where $\phi_{ij}$ is the relative angle between two particles in the $xy$-plane.
The RSPDM $\rho_{CP}({\bf r},{\bf r}',t)$ of the wavefunction $\psi_{CP}$ is uniquely defined, and it can be used to obtain one-body observables, by properly accounting the gauge. 
For example, the Fourier transform of $\rho_{CP}({\bf r},{\bf r}',t)$ yields the canonical rather than the kinetic momentum distribution because of the presence of
vector potential interactions.
In order to obtain the kinetic momentum distribution, one should first find 
a basis of eigenstates of the kinetic momentum operators.
However, this is not possible because the $x$ and $y$ components of these operators do not commute at the positions of the particles where the flux is present: 
\begin{equation}
[p_{x,i}^a,p_{y,i}^a]=-i 2 \pi \alpha\sum_{j \neq i}  \delta({\bf r}_i{-}{\bf r}_j),
\end{equation}
where $p_{x,i}^a{\bf\hat x} {+} p_{y,i}^a {\bf\hat y}
\equiv {-}i {\bf \nabla}_i {+} \alpha \sum_{j \neq i}  {\bf \hat z} {\times} {\bf r}_{ij}   / r_{ij}^2$.
Therefore, unlike the case for bosons or fermions, the kinetic momentum distribution for
anyons is not a proper observable.
In order to remedy this situation, we study the expansion of anyons from an initially localized state, to find an appropriate observable that corresponds to the momentum distribution, which reduces to the usual definitions when the statistical parameter $\alpha$ approaches $0$ for bosons or $1$ for fermions.

For clarity, we first discuss free expansion of two anyons released from a harmonic trap. When $N{=}2$, the Schr\"odinger equation
$i \frac{\partial}{\partial t}\psi{=}H\psi$
can be rewritten in center-of-mass ${\bf R}{=}\left( {\bf r}_1{+}{\bf r}_2 \right)/2 {\equiv} \left(R, \theta \right)$ and the relative ${\bf r}{=}{\bf r}_1 {-}{\bf r}_2 {\equiv} \left(r,\phi \right)$
coordinates.
The ground state for two anyons in a harmonic potential is given by~\cite{Wilczek1982,Leinaas1977,Wu1984},
\begin{equation}
	\psi 	({\bf R},{\bf r},t=0) = {\mathcal N}_2 r^{\left|\alpha\right|} e^{i \alpha \phi} e^{-R^2-\frac{r^2}{4}},
\label{phi2}
\end{equation}
where ${\mathcal N}_2$ is the normalization constant.
Equation \eqref{phi2} already shows two important characteristics of fractional statistics ($0{<}{\left|\alpha\right|}{<}1$), with all their implications: the wavefunction cannot be written as a product of single-particle wavefunctions, and it is not single-valued.
At $t{=}0$, the trap is turned off and two anyons start expanding.
The expansion dynamics can be found by decomposing the wavefunction \eqref{phi2} into two-anyon eigenstates in free space, which are given by~\cite{Aharonov1959}
\begin{equation}
	\phi_{KkMm} ({\bf R}, {\bf r}) = e^{i M \theta} J_{\left|M\right|}(K R) e^{i (m+\alpha) \phi} J_{\left| m + \alpha \right|}(k r),
	\label{phik2}
\end{equation}
up to normalization, with the corresponding energy $E_{Kk} {=} K^2/4{+}k^2$.
The principal quantum numbers are $\{K,k\} {\in} [0, \infty \rangle$, and the angular quantum numbers are $\{M,m\} {\in} \mathbb{Z}$.
Because the initial ground state is rotationally invariant, only eigenstates with $M{=}m{=}0$ are present in the expansion; therefore, we omit $M$ and $m$ in further notation.
The time-dependent wavefunction during free expansion of two anyons is
\begin{align}
	\label{psi2}
	& \psi 	({\bf R},{\bf r},t>0)
	= \int{dK dk K k}
	a_{Kk} \phi_{Kk} e^{-i E_{Kk} t}\\
	&\propto \frac{1}{(1+it)^2}\left(\frac{r}{1+it}\right)^{\left|\alpha\right|}
	{\exp}\left[-\frac{R^2+r^2/4}{1+it} + i \alpha \phi \right], \nonumber
\end{align}
where the coefficients $a_{Kk}$ are the projection coefficients of the initial wavefunction~\eqref{phi2} on eigenstates in free space~\eqref{phik2}:
\begin{equation}
	\label{ak}
	a_{Kk}
 \propto k^{\left|\alpha\right|} e^{-\frac{K^2}{4}-k^2}.
\end{equation}
We identify $|a_{Kk}|^2$ with the quasimomentum distribution of two anyons.

That this definition is natural is underpinned by the following observations:
(i) the quasimomentum distribution does not change during free expansion;
(ii) this definition reduces to the standard one when the statistical parameter $\alpha$ approaches $0$ for bosons or $1$ for fermions;
(iii) the asymptotic form of the single-particle density $\rho({\bf r},t{\rightarrow}\infty)$ has the same shape as $|a_{Kk}|^2.$
Observation (i) is evidently true, observation (ii) follows from the fact that eigenstates for bosons and fermions in free space are built from plane waves (properly symmetrized), and we have verified (iii) to hold explicitly.
The generalization of Eq. (\ref{psi2}) to the case of $N$ anyons would read $\psi(t{>}0)
=\int{d\beta} a_{\beta} \phi_{\beta} e^{-i E_{\beta} t}$.
However, a definition of the single-particle quasimomentum distribution from the projection coefficients $a_{\beta}$ is unclear, as we do not know which quantum numbers $\beta$, define eigenstates $\phi_{\beta}$ of $N$ anyons in free space.
These eigenstates are complex many-body wavefunctions, because a system of anyons is a genuine many-body problem with all its inherent difficulties, even though Hamiltonian $H$ appears as to describe noninteracting particles. The fact is that vector potential interactions between particles from $H_{CP}$, when gauged out to obtain $H$, remain hidden in the anyonic symmetry of the wavefunction $\psi$.
Nevertheless, we can {\it define} the quasimomentum distribution for $N$ anyons as the asymptotic single-particle density, after expansion in free space. This definition obviously obeys observation (ii) above, although the connection with projection coefficients $a_{\beta}$ is not yet clear.

\begin{figure}
	\centerline{\mbox{\includegraphics[width=0.35\textwidth]{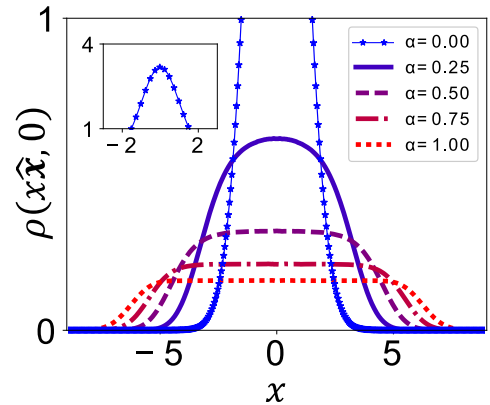}}}
	\caption{
		(color online) The cross section of the single-particle density, $\rho(x{\bf \hat x},0)$, of the wavefunction \eqref{psit}, which we identify with quasimomentum distribution (see text), for $N=20$ and different values of $\alpha$, obtained via Monte-Carlo integration. The inset shows the peak of the bosonic density at $\alpha=0$.
	}
	\label{qmfig}
\end{figure}

\begin{figure*}[htbp]
	\centerline{\mbox{\includegraphics[width=1.0\textwidth]{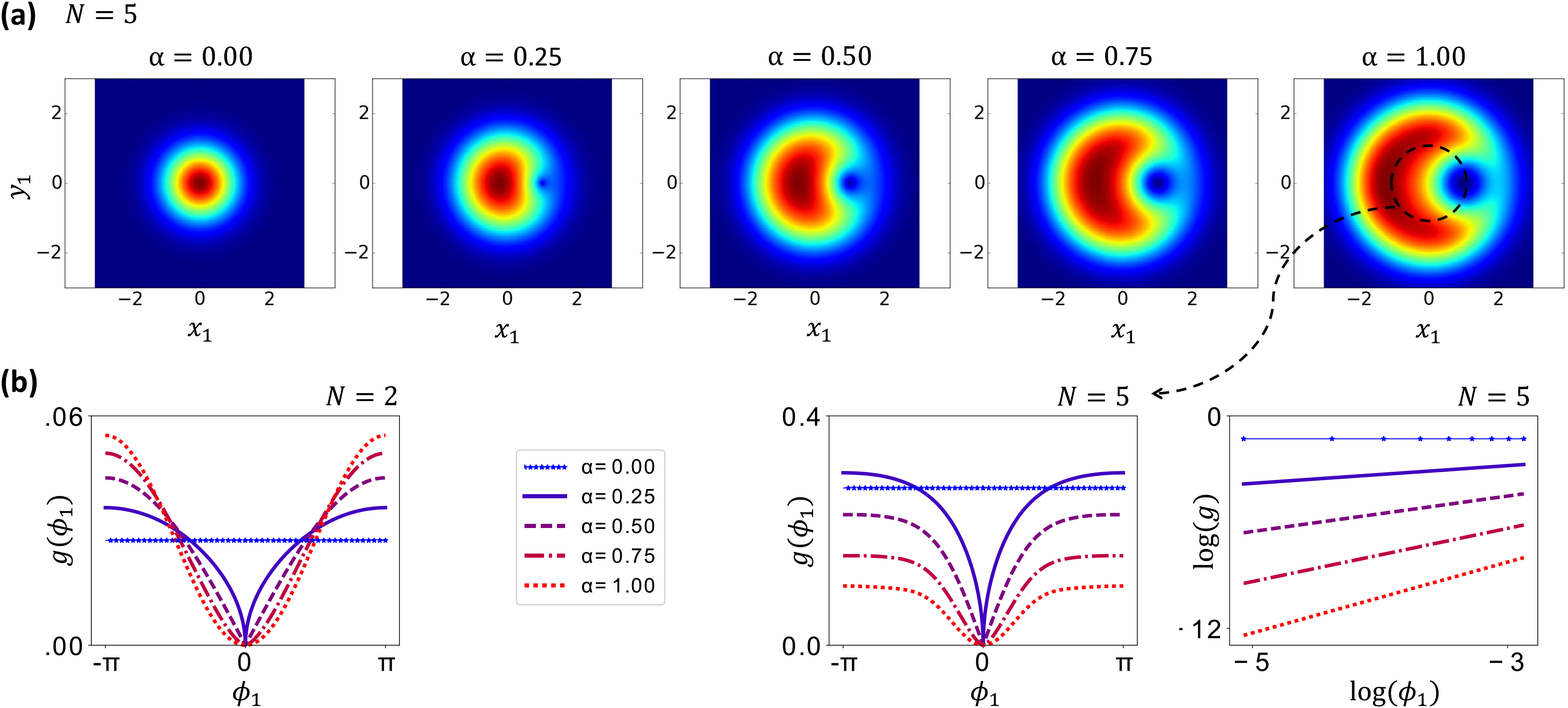}}}
	\caption{
		(color online)
		Pair correlations of anyons.
		(a) Pair correlation function $g({\bf r}_1,{\bf r}_2{=}{\bf \hat x})$ for different values of the statistical parameters $\alpha$.
		(b) Dependence of $g$ on the distance between particles $|{\bf r}_1-{\bf r}_2|=|(\cos\phi_1-1) {\bf \hat x} + \sin\phi_1 {\bf \hat y}|$ as a function of $N$ and $\alpha$. See text for details.}
	\label{pairfig}
\end{figure*}

Let us now consider expansion of $N$ anyons from the harmonic trap.
The generalization of wavefunction~\eqref{phi2} for $N{>}2$ does not yield the ground state in a harmonic oscillator.  In order to gain understanding of the expansion dynamics of anyons, we assume that initially the system is in its eigenstate, given by~\cite{Wu1984}
\begin{equation}
\psi (\{ {\bf r}_i \},t{=}0)
={\mathcal N}_N\prod_{i<j} r_{ij}^{{\left|\alpha\right|}} e^{i \alpha \phi_{ij}} e^{-\sum_{k=1}^N \frac{|{\bf r}_k|^2}{2}}.
\end{equation}
We can obtain the dynamics of the system for $t{\geq}0$ by employing the scaling transformation~\cite{Lewis1969,Minguzzi2005}:
\begin{equation}
\label{psit}
\psi(\{ {\bf r}_i \},t{>}0) =\frac{1}{b^N}
\psi(\{\frac{{\bf r}_i}{b}\},0)
e^{i\frac{\dot b}{2 b } \sum_{k}^{N}{|{\bf r}_k|^2}}
e^{-i E_N \tau(t)} .
\end{equation}
Here $b{=}\sqrt{1+t^2}$ is the time dependent scaling factor, $E_N{=}N{+}\alpha{N(N{-}1){/}2}$ the eigenstate energy and $\tau(t){=}\int^t\frac{dt'}{b^2(t')}$ a scaled time.
The evolution of the single-particle density is self-similar. Consequently, the shape of the asymptotic single-particle density is the same as the initial single-particle density. In Fig. \ref{qmfig} we plot its cross section for $N=20$, for different values of the statistical parameter $\alpha$. We see that on the bosonic side, the form is narrower and sharper, and it gets flatter and broader on the fermionic side, as one would expect from the quasimomentum distribution.

Another observable that depends on the statistical parameter $\alpha$, and can be used to obtain information on the anyonic character of the system is the pair correlation function~\cite{Zhang2014}. The pair correlation function
$g({\bf r}_1,{\bf r}_2,t)=N(N-1)\int{|\psi({\bf r}_1,{\bf r}_2,\ldots,{\bf r}_N,t)}|^2 d{{\bf r}_3}...d{{\bf r}_N}$ is illustrated in Fig.~\ref{pairfig} for different values of $\alpha$ and $N$, at $t{=}0$ (it changes trivially with time due to the self-similarity of the evolution). One particle is fixed at ${\bf r}_2{=}{\bf \hat x}$, and we show $g$ as a function of the position of the second particle, ${\bf r}_1$.
In Fig.~\ref{pairfig}(a) we show $g$ for $N{=}5$: two particles are uncorrelated only in the bosonic limit ($\alpha=0$), coinciding with the case with no repulsive statistical interactions at any distances~\cite{Huang1995,Mancarella2012}.
In Fig. ~\ref{pairfig}(b), the angle $\phi_1$ parametrizes the position of the moving particle, ${\bf r}_1=\cos\phi_1 {\bf \hat x} + \sin\phi_1 {\bf \hat y}$.
Suppose that we perform expansion of two anyons from a harmonic trap. If we detect one anyon at an angle $\phi_2=0$, one may ask what is the probability of detecting the second anyon at some other angle $\phi_1$?
The plot in Fig.~\ref{pairfig}(b) for $N=2$ provides information on what may we expect from
such an experiment. Bosons would be completely uncorrelated with probability independent of $\phi_1$,
fermions anti-correlated with the peak of the probability at $\phi_1=\pi$, and anyons
ranging in between these two cases, depending on $\alpha$. From the structure of the wavefunction \eqref{psit} it follows that for small interparticle distances, the pair correlation function scales as
$g\propto|{\bf r}_1-{\bf r}_2|^{2{\left|\alpha\right|}}$; this can be seen from the plot in Fig.~\ref{pairfig}(b).
It was recently pointed out that in topologically ordered states, the low energy onset of spectral functions scales as a power law with the statistical parameter in the exponent~\cite{Morampudi2017}.

One possible route to implement free expansion of anyons in ultracold atomic gases is by building
upon the proposals in Refs.~\cite{Paredes2001,Zhang2014,Lundholm2016}.
Suppose that one implements a system proposed in Ref.~\cite{Paredes2001},
where one first introduces a system of hard-core bosons in a synthetic magnetic field, which
is represented by a bosonic version of the Laughlin wavefunction for electrons in the
quantum Hall regime~\cite{Paredes2001}. Next, instead of creating quasihole fractionalized excitations
with lasers~\cite{Paredes2001}, suppose that one introduces a few bosonic atoms of another species, which have hard-core
repulsive interactions with the original bosons. These newly introduced bosons would behave as
anyons, and their expansion in the presence of background of the original bosons
would reveal the quasimomentum distribution discussed here.
Such a system of test particles immersed in a fractional quantum Hall like state was discussed in Refs.~\cite{Zhang2014,Lundholm2016}.

In conclusion, we have shown that the momentum distribution, which played a key role in demonstrating Bose-Einstein condensation~\cite{Cornell2002} and Fermi degeneracy~\cite{DeMarco1999}, is not a proper observable for a system of anyons. 
Instead, we pointed out that the asymptotic single-particle density following the expansion of anyons from an initially localized state has all the characteristics of the quasimomentum distribution, and it reduces to the standard definitions when the statistical parameter approaches $0$ for bosons and $1$ for fermions.
We have obtained the quasimomentum distribution for $N$ anyons expanding from an eigenstate in a harmonic trap.
We demonstrated that two-particle correlations of this state scale as a power-law with the statistical parameter in the exponent. Finally, we have proposed a possible implementation of this system with ultracold atomic gases.
Our work points out at intriguing aspects associated with extracting observables from anyonic wavefunctions that are still awaiting to be explored. An interesting problem to consider next is expansion of non-Abelian anyons. 

We acknowledge useful discussions with M. Solja\v{c}i\'{c}, A. Trombettoni, P. Calabrese, J. Goold, and B. Bu\v{c}a.
This work was supported by the Croatian Science Foundation grant IP-2016-06-5885 SynthMagIA, and in part by the QuantiXLie Center of Excellence.


\end{document}